# Imaging of Spatial Correlations of Two-Photon States

*Ivan B. Bobrov[1,2], Dmitry A. Kalashnikov[1], and Leonid A. Krivitsky[1,*]*

[1]Data Storage Institute, Agency for Science, Technology and Research (A*STAR), 117608 Singapore

[2] Department of Physics, M. V. Lomonosov Moscow State University, 119991 Moscow, Russia

*e-mail: *Leonid_Krivitskiy@dsi.a-star.edu.sg*

We use a fiber based double slit Young interferometer for studying the far-field spatial distribution of the two-photon coincidence rate (coincidence pattern) for various quantum states with different degree of spatial entanglement. The realized experimental approach allows to characterize coincidence patterns for different states without any modifications of the setup. Measurements were carried out with path-entangled and separable states. The dependence of the coincidence pattern on the phase of the interferometer for superposition and separable states was studied. The results have implications for using of nonclassical light in multiphoton imaging, quantum lithography, and studies of phase decoherence.

**Introduction.**

Nonclassical states of light have great promise to advance numerous optical technologies, including optical sensing [1, 2], spectroscopy [3-5] and imaging [6-12]. It has been shown theoretically [3, 4] and experimentally [5] that the rate of two-photon absorption can be enhanced by light with nonclassical frequency correlations. An analogous enhancement using spatially entangled photons is also suggested in the context of quantum lithography [7]. The implementation of *N*-photon entangled states allows surpassing the diffraction limit by the factor of *N* [8-18]. The technique holds a promise to yield much smaller feature size of the structured material, compared to the lithography with the classical light. However, several theoretical works have raised doubt about the possibility of the enhancement effect with such states and even suggested that the spatial entanglement can only reduce the multiphoton absorption rate [19-23].

In a typical setting for the quantum lithography experiment, a $N$-photon entangled state is emitted from a double-slit of a Young interferometer. Spatial far field correlations are characterized by $N$ single photon detectors or CCD cameras with their signals addressed to a correlator [10]. An alternative approach, based on using an optical centroid technique, was suggested in Ref. [24] and experimentally realized in [11]. However, the experimental analysis in [11] was restricted to the study of a fixed two-photon entangled state.

A detailed study of spatial correlations of various two-photon entangled states was performed in [9]. The results of this work were used to elaborate the trade-off between the resolution and the efficiency of the quantum lithography [23]. A similar experimental approach has been exploited for studying spatial correlation of four-photon entangled states [25]. However, this approach is based on using an imaging system, and the transfer from one state to another requires inclusion of additional optical elements, and adjustment of their parameters [9, 25].

In this work we implement a fiber based double slit Young interferometer with phase stabilization, which allows revealing the far-field correlations of arbitrary two-photon states with different degree of spatial entanglement. Different quantum states are characterized at absolutely the same settings of the setup. The approach also allows simple and accessible preparation of various quantum states, making it an ideal test-bed for experiments in quantum state engineering and quantum imaging.

**Theory**

Let us consider the arrangement of the Young double slit interferometer, where the two slits are formed by the outputs of two single mode fibers, see Fig.1. The quantum state is emerging from the tips. Two fibers are separated by the distance $d$, with the effective diameter of the Gaussian mode $R$. Fibers are aligned parallel to each other and their tips are located at the focal plane of the lens with the focal length $f \gg d$. The collimated beam is then split by a 50/50 beamsplitter and it is detected by two detectors, preceded by narrow slits with widths $a$ and coordinates $s_1$, $s_2$. The slits are scanned in the plane of the fiber tips, and coincidences of the detector photocounts are recorded.

Let us denote $\phi(\mathbf{u_1}, \mathbf{u_2})$ to be the two-photon wavefunction in the near field, where $\mathbf{u_j}$ is a two-dimensional position vector in the plane of the $j$-th ($j = 1, 2$) fiber tip. After the Fourier transformation, performed by the lens, the wavefunction in the far-field is given by

$$\psi(\mathbf{r_1}, \mathbf{r_2}) = \frac{1}{-\lambda^2 f^2} \int_{-\infty}^{\infty} d^2\mathbf{u_1}\, d^2\mathbf{u_2}\, \phi(\mathbf{u_1}, \mathbf{u_2}) exp\left[-\frac{2\pi i}{\lambda f}(\mathbf{r_1}\mathbf{u_1} + \mathbf{r_2}\mathbf{u_2})\right], \qquad (1)$$

where $\lambda$ is the wavelength of light. The two-photon coincidences rate, measured by two detectors with the slits at arbitrary positions $s_1$ and $s_2$, is given by [20]

$$P(s_1, s_2) = \int_{s_1-a/2}^{s_1+a/2} dx_1 \int_{s_2-a/2}^{s_2+a/2} dx_2 \int dy_1 \int dy_2 \, |\psi(\mathbf{r}_1, \mathbf{r}_2)|^2. \quad (2)$$

Assuming that the slit width is much smaller than the typical feature size of $|\psi(\mathbf{r}_1, \mathbf{r}_2)|^2$, we obtain

$$P(s_1, s_2) \approx \int dy_1 \int dy_2 \, |\psi(\mathbf{r}_1 = s_1\hat{\mathbf{x}} + t_1\hat{\mathbf{y}}, \mathbf{r}_2 = s_2\hat{\mathbf{x}} + t_2\hat{\mathbf{y}})|^2, \quad (3)$$

and $P(s_1, s_2)$ would be sampling of $|\psi(\mathbf{r}_1, \mathbf{r}_2)|^2$ along the $x$ direction.

Let us now consider particular examples of two-photon states at the outputs of the fibers. First, we consider the path entangled two-photon state, referred here as *NOON*-state

$$|\varphi_{NOON}\rangle = \tfrac{1}{\sqrt{2}}(|2,0\rangle + e^{i2\theta}|0,2\rangle) = \tfrac{1}{2}(a^{+2} + e^{i2\theta}b^{+2})|0\rangle, \quad (4)$$

where $|m, n\rangle$ denotes the Fock state with $m$, and $n$ photons in two spatial modes (outputs of the fibers), and $\theta$ is a relative phase. Following [20] we define the annihilation operator in the near-field as $A(\mathbf{u})$ with the communication relation $[A(\mathbf{u}), A^\dagger(\mathbf{u}')] = \delta(\mathbf{u} - \mathbf{u}')$. The operators $a^\dagger$ and $b^\dagger$ can be rewritten as

$$a^\dagger = \int d^2\mathbf{u}\, \Pi\left(\mathbf{u} + \tfrac{d}{2}\hat{\mathbf{x}}\right) A^\dagger(\mathbf{u}), \quad (5a)$$

$$b^\dagger = \int d^2\mathbf{u}\, \Pi\left(\mathbf{u} - \tfrac{d}{2}\hat{\mathbf{x}}\right) A^\dagger(\mathbf{u}), \quad (5b)$$

where $\Pi(\mathbf{u})$ describes the optical Gaussian mode of each output fiber, given by

$$\Pi(\mathbf{u}) = \tfrac{1}{\sqrt{2\pi}R} exp\left(-\tfrac{|\mathbf{u}|^2}{4R^2}\right), \quad (6)$$

where it is assumed that $d \gg R$. From Eqs. (5a), (5b) we obtain the two-photon wavefunction for the *NOON*-state in the near field

$$\phi(\mathbf{u}_1, \mathbf{u}_2)_{NOON} = \tfrac{1}{\sqrt{2}}\left[\Pi\left(\mathbf{u}_1 + \tfrac{d}{2}\hat{\mathbf{x}}\right)\Pi\left(\mathbf{u}_2 + \tfrac{d}{2}\hat{\mathbf{x}}\right) + e^{i2\theta}\Pi\left(\mathbf{u}_1 - \tfrac{d}{2}\hat{\mathbf{x}}\right)\Pi\left(\mathbf{u}_2 - \tfrac{d}{2}\hat{\mathbf{x}}\right)\right]. \quad (7a)$$

Using (1) we obtain the wavefunction in the far field

$$|\psi(\mathbf{r}_1, \mathbf{r}_2)|^2_{NOON} = \tfrac{1}{(2\pi\sigma^2)^2} exp\left[-\tfrac{1}{2\sigma^2}(x_1^2 + y_1^2 + x_2^2 + y_2^2)\right]\left\{1 + cos\left[\tfrac{2\pi}{\Lambda}(x_1 + x_2) - 2\theta\right]\right\}, \quad (7b)$$

where $\sigma \equiv \tfrac{\lambda f}{4\pi R}$ is the size of the localization of the interference pattern, and $\Lambda \equiv \tfrac{\lambda f}{d}$ is the period of interference fringes along $(x_1 + x_2)$ axis.

Let us now analyze the dependence of the interference pattern on the relative phase $\theta$. From Eq. (7b) it follows that the behavior of $|\psi(\mathbf{r}_1, \mathbf{r}_2)|^2_{NOON}$ is similar in two cases: (1) when the phase $\theta$ is fixed and the detectors are scanned along $x_1 = x_2$ direction, and (2) when positions of the detectors $\mathbf{r}_1, \mathbf{r}_2$ are fixed and the phase $\theta$ is scanned. These two configurations were realized in our experiment. Displacements of the fringes depending on the phase are schematically shown by red arrows in Fig. 2a.

Configuration when the two detectors are located at exactly the same positions, i.e. $\mathbf{r_1} = \mathbf{r_2}$, is equivalent to the detection of the two-photon spatial distribution by an array of two-photon absorbing *pixels* (or two-photon absorbing media). Without loss of generality we can assume $x_1 = x_2 = 0$ and $y_1 = y_2 = 0$, so that Eq. (7b) becomes

$$|\psi(\mathbf{0},\mathbf{0})|^2_{NOON} = \frac{1}{(2\pi\sigma^2)^2}(1 + cos[2\theta]). \tag{7c}$$

From Eq. (7c) it follows that the two-photon absorption rate in each *pixel* can be effectively modulated with unity visibility solely by driving the phase $\theta$.

Let us now consider the case of a spatially separable two-photon state,

$$|\varphi_{Sep}>= |1,1\rangle = (a^+ b^+)|0\rangle. \tag{8}$$

Similar to (7a) we obtain the wavefunction in the near field

$$\phi(\mathbf{u_1}, \mathbf{u_2})_{Sep} = \frac{1}{\sqrt{2}} e^{i\theta}\left[\Pi\left(\mathbf{u_1} + \frac{d}{2}\hat{x}\right)\Pi\left(\mathbf{u_2} - \frac{d}{2}\hat{x}\right) + \Pi\left(\mathbf{u_1} - \frac{d}{2}\hat{x}\right)\Pi\left(\mathbf{u_2} + \frac{d}{2}\hat{x}\right)\right]. \tag{9a}$$

The resulting wavefunction in the far field is given by

$$|\psi(\mathbf{r_1}, \mathbf{r_2})|^2_{Sep} = \frac{1}{(2\pi\sigma^2)^2} \exp\left[-\frac{1}{2\sigma^2}(x_1^2 + y_1^2 + x_2^2 + y_2^2)\right]\left\{1 + cos\left[\frac{2\pi}{\Lambda}(x_1 - x_2)\right]\right\}. \tag{9b}$$

From (9b) it follows that the interference fringes in the far field are observed along $(x_1 - x_2)$ axis. Note that unlike for the case of the spatially-entangled *NOON*-state, $|\psi(\mathbf{r_1}, \mathbf{r_2})|^2_{Sep}$ is phase insensitive.

The resulting spatial distributions for the two states, given by Eqs. (7b) and (9b), are shown in Fig. 2(a, b). The parameters of the calculations correspond to the ones used in the experiment, see below. In the performed experiment the patterns are scanned in two directions: along $x_1 = x_2$ axis and along $x_1 = -x_2$ axis, which are shown in Fig. 2(a, b) by dashed lines. Note that for the both cases considered above (*NOON* and separable states), the states are still entangled in the frequency and temporal degrees of freedom.

**Experiment**

In the experiment the two-photon states are produced via spontaneous parametric down conversion (SPDC). A beam of a continuous wave (CW) diode laser at 407 nm is focused in a 2 mm long type-II β-barium borate crystal (BBO1) by a lens (L1) with focal length *f*=300 mm, see Fig.3. The crystal is cut for type-II SPDC in collinear frequency degenerate regime. A UV-Mirror (UVM) filters the pump after BBO1 and fluently transmits the SPDC. To compensate for spatial and temporal walk-offs another 1mm long BBO crystal (BBO2) with its optical axis parallel to the axis of the BBO1 and a half wave plate (HWP2) at 45° are placed after the UVM [26].

The SPDC with the central wavelength at 814 nm is split by a polarization beam splitter (PBS1) and coupled into two single mode fibers SMF1, SMF2 (HP-780) by objective lenses O1, O2 with $f$=15.3 mm, placed at the distances of 600 mm from the BBO1 [27]. A half wave plate HWP3 placed in front of the PBS1 controls the state at the input of the two fibers, given by

$$|\varphi\rangle = \frac{\sin(4\alpha)}{\sqrt{2}}(|2,0\rangle + |0,2\rangle) + \cos(4\alpha)|1,1\rangle =$$
$$= \left[\frac{\sin(4\alpha)}{2}(a^{\dagger 2} + b^{\dagger 2}) + \cos(4\alpha)a^\dagger b^\dagger\right]|0,0\rangle, \qquad (10)$$

where $\alpha$ is the orientation of the HWP3. When $\alpha = 0$ we obtain the separable state, and when $\alpha = \pi/8$ we obtain the *NOON*-state.

Bare outputs of SMF1, and SMF2 are fused together in a microfiber array (Chiral Photonics) with the pitch size $d$=72 μm. The diameter of the effective mode of the SMF is $R$=4.3 μm. The light from both fibers is collimated by a lens L2 with $f$=60 mm. The array is tilted to produce interference fringes in the vertical plane (the fiber tips are oriented in the horizontal plane). The relative phase between the fiber outputs is controlled by a retro-reflecting trombone prism (TP), mounted on a motorized stage (Owis, resolution 500 nm), and by a mirror M4, mounted on a piezo-driven stage (Melles Griot, resolution 20 nm). A beam of a femtosecond laser with pulse duration 84 fs is used for alignment of the optical paths: the visibility of the classical interference pattern from the fs-laser is optimized depending on the position of the TP. The phase between the channels of the interferometer is locked by using an auxiliary CW He-Ne laser at 632 nm, which is launched along the path of the SPDC. The interference pattern from the He-Ne laser is reflected from the interference filter (IF), and observed by the CCD. The phase is controlled by changing the voltage of the piezo-translator of M4.

The polarization of the SPDC at the outputs of the array is controlled by quarter- and half-wave plates (QWP1, QWP2 and HWP4, HWP5, respectively), so that the SPDC transmission through a polarization beamsplitter PBS2 is maximized. The SPDC is filtered by an IF with a central wavelength of 814 nm, and the full-width on the half maximum of 20 nm.

In the measurement part of the setup the SPDC is split by a 50/50 beamsplitter (NPBS), and the coincidence pattern is scanned by two slit detectors, placed at reflected and transmitted arms. Slit detectors (referred further to as detectors) are realized using tips of multimode fibers MMF1, and MMF2 with a core diameter $a$=62.5 μm, preceded by cylindrical lenses CL1, and CL2 with $f$=70 mm and $f$=80 mm which focus light in the vertical plane. The cylindrical lenses are used to enhance the collection efficiency and do not influence the observed coincidence distributions. Detectors are mounted on motorized translation stages (Thorlabs, resolution 1μm). The outputs of multimode fibers are connected to single photon avalanche photodiodes (APD, Perkin-Elmer AQRH-14FC). The signals from the APDs are addressed to a coincidence circuit (Ortec-567 TAC/SCA, time window 7 ns).

The generated states are characterized by directly analyzing coincidences between photons emerging from the outputs of the interferometer. This is done by erecting a flipping mirror (FM) after the array, and by using a combination of a HWP6 and a PBS3 as a beamsplitter. Adjustable objective lenses O3, and O4 with $f$=40 mm are used to couple light into multimode fibers MMF3, and MMF4, which are connected to the APDs.

**Results and discussion**

First, we characterize the generated states. We erect the FM, and adjust objective lenses O3, and O4 so that both MMF3, and MMF4, receive the SPDC from a single output fiber of the interferometer. The dependence of the coincidence count rate between the APDs on orientation of the HWP3, is shown in Fig. 4(a). According to (10) the separable state $|1,1\rangle$ is generated at $\alpha = 0 + \pi n/4, n \in \mathbb{Z}$, and the *NOON*-state is generated at $\alpha = \pi/8 + \pi n/4$. The experimental dependence exhibits minimum at $\alpha = 0$, and the maximum at $\alpha = \pi/8$ and yields 97.5±2 % visibility, which agrees with the theory. Next, the objective lens O3 is readjusted, so that MMF3 receive the SPDC from a different output fiber of the interferometer. The corresponding dependence of the coincidence count rate is shown in Fig. 4(b). The maximum is observed at $\alpha = 0 + \pi n/4$, which corresponds to generation of the separable state, and the minimum is observed at $\alpha = \pi/8 + \pi n/4$, which corresponds to generation of the *NOON*-state. The dependence yields 95±3 % visibility. The observed trends and visibilities of the coincidence dependencies ensure high fidelity of generated states.

Then, we proceed with the study of far field two-photon coincidence patterns. First, we measure the pattern for the *NOON*-state state. We fix the orientation of the HWP3 at $\alpha = \pi/8$, and set the phase to $\theta$=0. The dependence of the coincidence count rate on the relative displacement of the two detectors in the same direction ($x_1 = x_2$) is shown in Fig. 5(a) and for different directions ($x_1 = -x_2$) in Fig. 5(b). The sinusoidal fringes (visibility 72±5.5 %) are observed only in one direction ($x_1 = x_2$), while in another direction ($x_1 = -x_2$) the smooth envelope of the spatial distribution is observed. The result agrees with the theoretical predictions in Fig. 2(a). Note that the maximum visibility of the two-photon interference for the classical light is 50%. The observed value of the visibility in our experiment is about 4 standard deviations higher than the classical limit.

Next, we fix the orientation of the HWP3 at $\alpha = 0$, and measure the coincidence pattern for the separable state. The results for relative displacement of the two detectors in the same direction ($x_1 = x_2$) are shown in Fig. 6(a), and for different directions ($x_1 = -x_2$) in Fig. 6(b). Unlike for the case of the *NOON*-state, the former measurement reveals the smooth envelope, and the latter reveals the fringes (visibility 84±3 %). The results are consistent with the theoretical predictions in Fig. 2(b).

For both experiments we found the period of the fringes to be $\Lambda$ = 566 ± 14 µm. The obtained value is about 16% less than the theoretical value of 678 µm. The discrepancy can be explained by the mismatch between the position of the tip and the exact focal point of the lens L2, which leads to slight focusing of the beam.

To study the phase dependence of the coincidence patterns we fix the position of pinhole detectors at $x_1 = x_2 = 0$ and change the phase by controlling the voltage on the piezo-stage of M4. Such configuration is sufficient to demonstrate the basic effect of the phase detuning on the modification of the fringes. In accordance with Eq. (7c), the *NOON*-state should demonstrate the sinusoidal dependence on the phase with the period equal to $\pi$. This was confirmed in the experiment, see Fig. 7(a). The visibility of the dependence is 77±4.5 %. Non unity visibility is most likely caused by high-frequency phase oscillations, which are not compensated by the feedback mechanism of the interferometer. Thus we experimentally confirm that the phase detuning in the case of the *NOON*-state results in the translation of interference fringes along $x_1 = x_2$ direction, see Fig. 2(a).

Strong dependence of the observed pattern on the relative phase $\theta$ opens a practical way of studying phase decoherence processes with spatially-entangled *NOON*-states in quantum imaging experiments [28]. Indeed, application of well-controlled phase modulation $\theta(t)$ (for example, by modulating the displacement of the piezo-driven mirror in our setup) would lead to controllable degradation of the interference visibility and corresponding modification to the interference pattern.

In contrast, the pattern for the separable state exhibits phase-invariant behavior, see Eq. (9b). The measured visibility of the dependence yields ~10 %. Slight discrepancies in the experimental values of visibilities are likely to result from imperfections in alignment of the measurement system and in the control of the phase. This result indicates that in experiments with spatially separable states the stable interference pattern can be observed without the necessity of stabilization of the phase of the interferometer.

In the presented experiments the states are highly entangled in frequency and temporal degrees of freedom. Detuning of the interferometer from the position of the optimal temporal overlap would allow to study degradation of the interference visibility. Note, that the rotation of HWP3 in our experimental setup allows preparing mixtures of *NOON*-state and separable states see Eq. (10). Moreover it is possible to use combinations of type-I and type-II crystals which operate in a collinear degenerate SPDC for generation of arbitrary two-photon states at the output of the interferometer [29, 30]. Another interesting possibility is to use the setup for studies of coincidence patterns for multiphoton states with $N > 2$ [24].

**Conclusions**

In conclusions we experimentally implemented a fiber based Young interferometer, which allows observation of the two-photon coincidence pattern in the far field for a variety of quantum states. Our experiment allows simple and flexible transfer from different states by rotating polarization elements, and controlling the phase of the interferometer. The experimental results obtained for two types of states are in good agreement with theoretical predictions. The dependence of the observed pattern on the phase of the interferometer was studied for *NOON* and separable states. The approach can be implemented for studies of controllable phase decoherence of spatially-entangled *NOON*-states, and far field coincidence patterns of *N*-photon (*N*>2) quantum states.

**Acknowledgements**

We acknowledge Mankei Tsang and Sergei Kulik for fruitful discussions. I. B. is grateful for the support of his research stay in Singapore by the Singapore International Pre-Graduate Award (SIGPA).

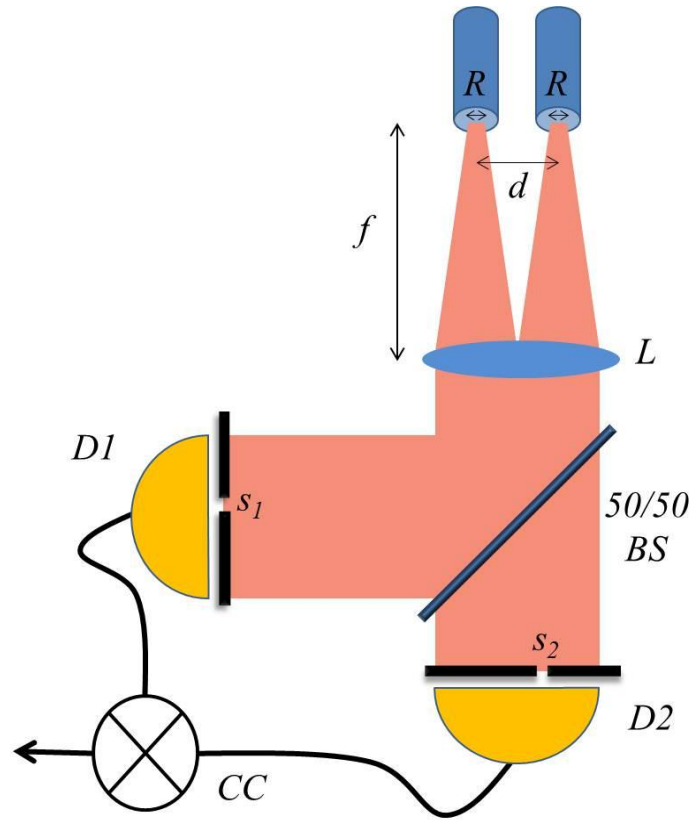

**Figure** 1 (Color online) The quantum state is emerging from the tips of two parallel single mode fibers with the effective diameter of the Gaussian mode $R$ and separated by the distance $d$. The fiber tips are located at the focal plane of the lens (L) with the focal length $f$. The collimated beam is split by a 50/50 beamsplitter (BS) and detected by two detectors (D1, D2), preceded by two slits with coordinates $s_1$, $s_2$. The slits are scanned in the plane of the fiber tips. The signals from the detectors are addressed to a coincidence circuit (CC).

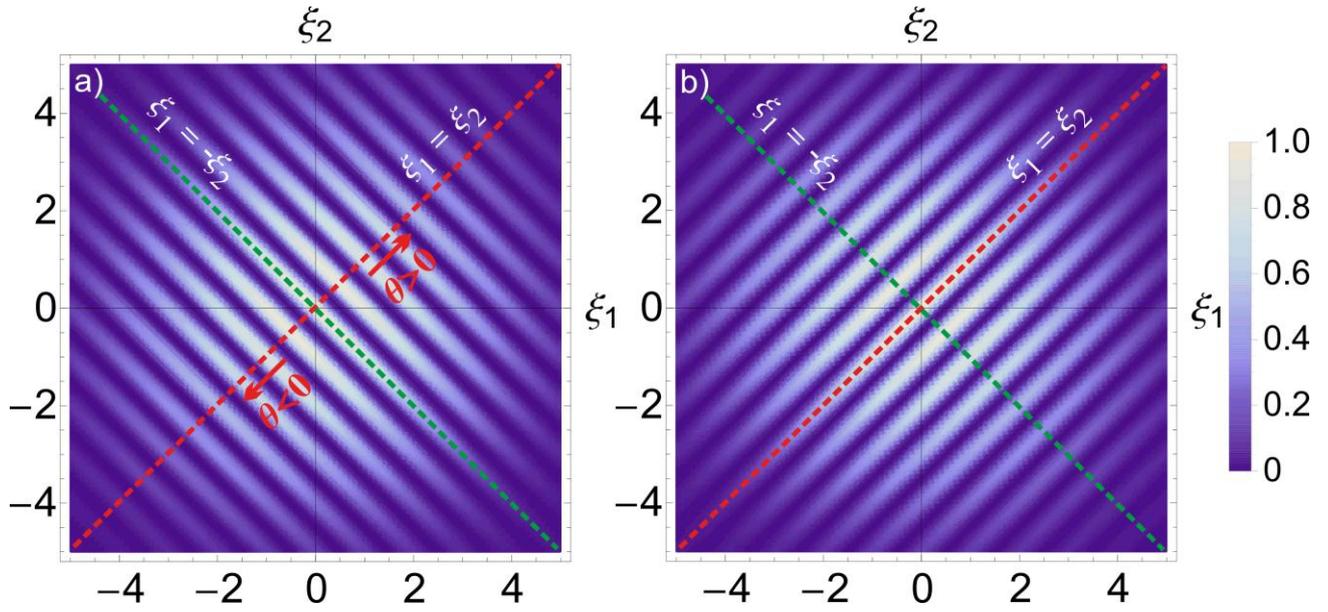

**Figure** 2 (Color online) Spatial distribution of coincidences calculated according to 7(b) and 9(b) for (a) *NOON* and (b) for separable states, respectively. For convenience the calculations are carried out with dimensionless parameters $\xi_i = \frac{x_i}{\Lambda}$, where $\Lambda = \frac{\lambda f}{d}$, $i = 1, 2$. Dashed lines show the directions of scanning of positions of slit detectors realized in the experiment, and red solid arrows in (a) schematically show displacement of the fringes, depending on the change of the phase $\theta$.

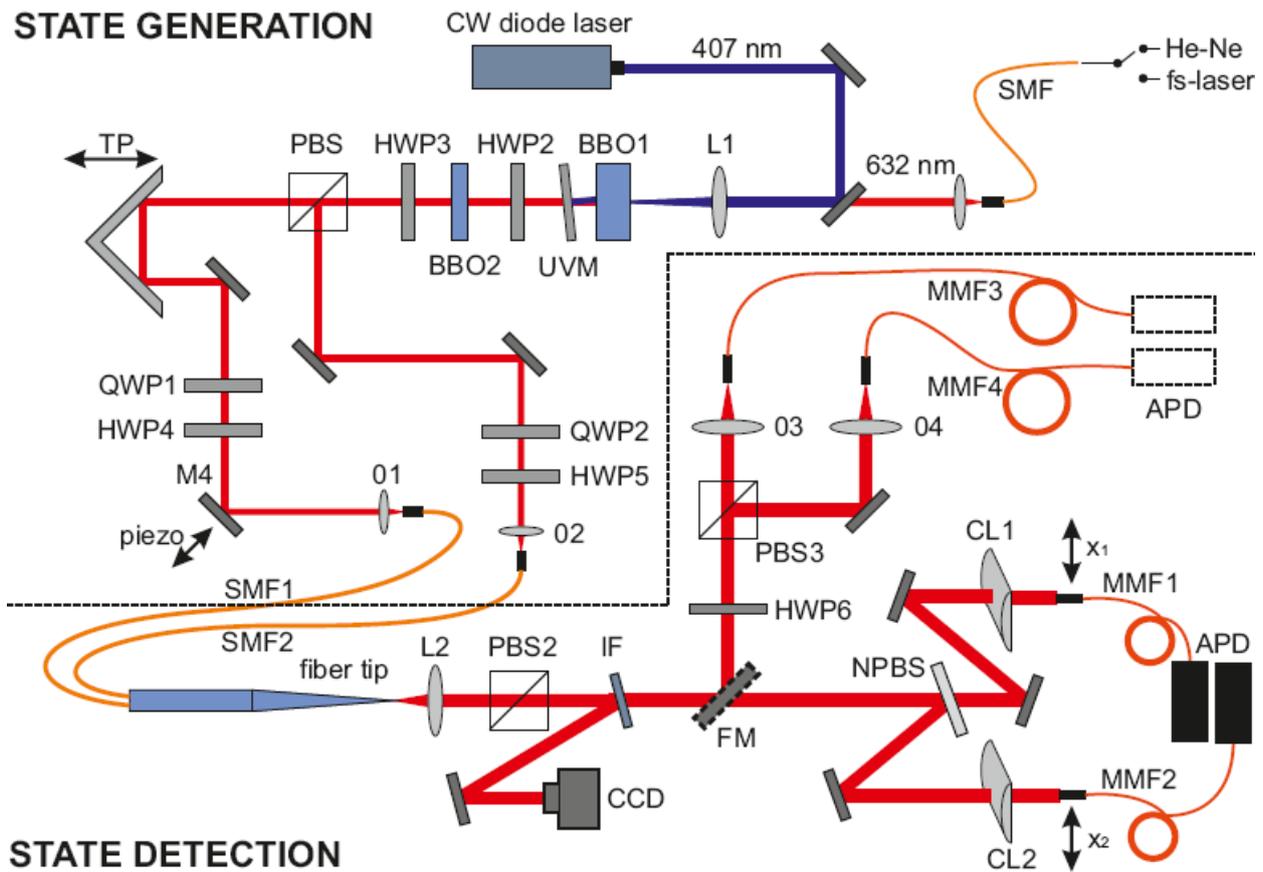

**Figure 3** (Color online) A diode laser pumps a BBO crystal (BBO1), where SPDC occurs. A UV-mirror (UVM) filters the pump. A half-wave plate (HWP2) and an additional BBO crystal (BBO2) compensate the walk-off. A half wave plate (HWP3) and a polarization beamsplitter (PBS1) are used for state preparation. The SPDC is coupled into two single mode fibers (SMF1,2) by objective lenses (O1,2). A trombone prism (TP), and a piezo-driven mirror (M4) are used to control the phase in the interferometer. Retardation plates (QWP1, 2 and HWP4, 5) and PBS2 control polarization at the outputs of SMFs. Fiber tips are fused in a microarray and positioned in a focal plane of a lens (L2). Interference filter (IF) reflects the phase locking beam to the CCD and transmits the SPDC. The SPDC is split by a 50/50 beamsplitter (NPBS), and collected by multimode fibers (MMF1, 2) preceded by cylindrical lenses (CL1, 2). The MMFs are plugged into avalanche photodiodes (APD) with their outputs addressed to a coincidence circuit (not shown). States are characterized by erecting a flipping mirror (FM), so that SPDC is coupled to multimode fibers MMF3, 4 with objective lenses O3, O4. A HWP6 at 45° and a PBS3 are used as a variable beamsplitter. A He-Ne laser is used for controlling the phase, and a femtosecond laser is used for alignment of the interferometer.

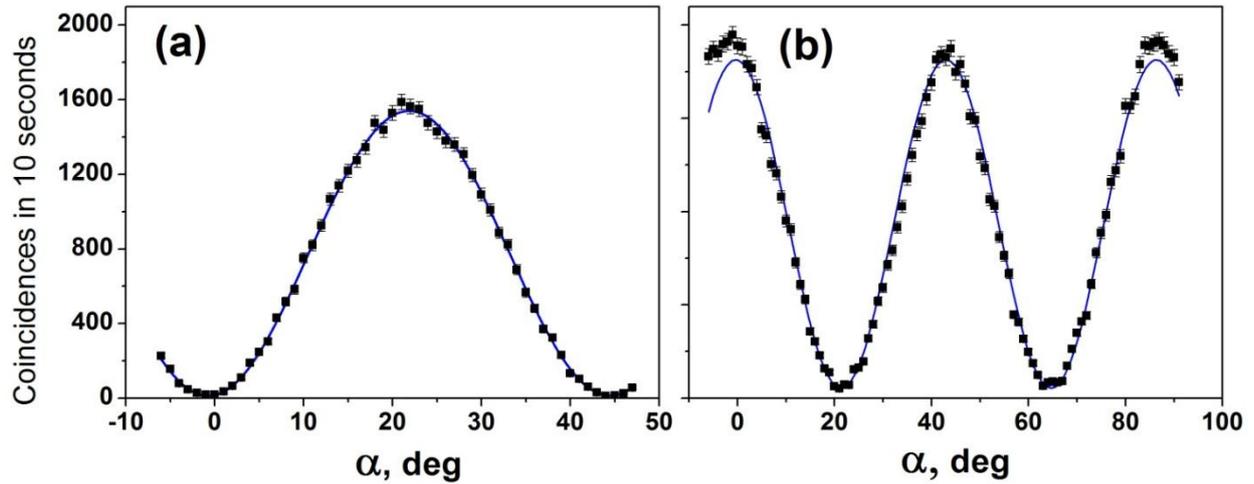

**Figure** 4 (Color online) Dependence of the coincidence count rate on the orientation angle $\alpha$ of HWP3, when the APDs detect SPDC from the same (a) and from different (b) outputs of the interferometer. Accidental coincidences (~100 coincidences per 10 sec) are subtracted. Solid lines are theoretical dependencies.

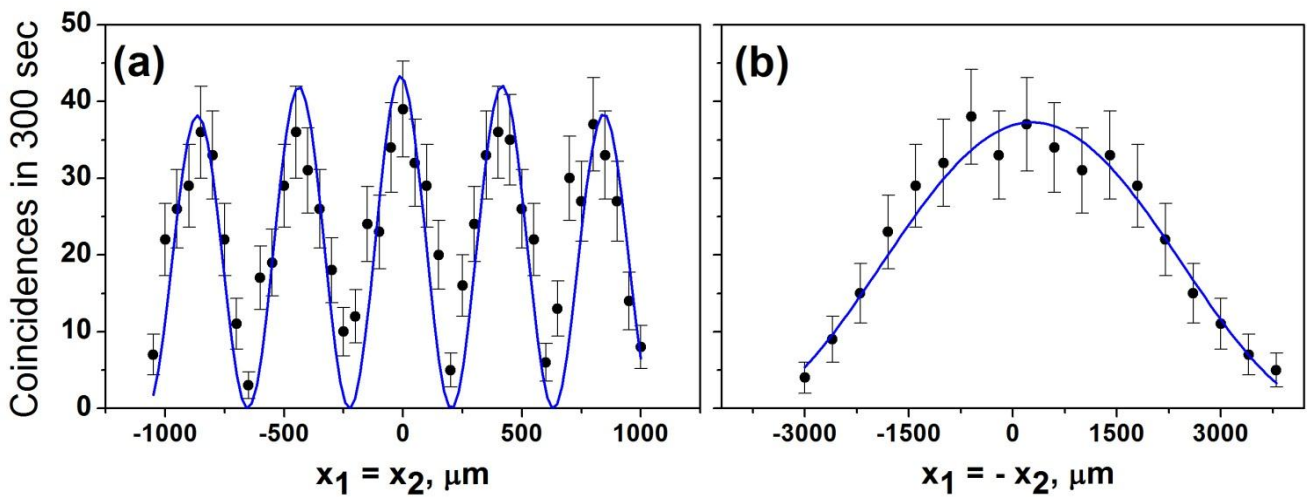

**Figure 5** (Color online) Dependence of the coincidence count rate for *NOON*-state on the relative displacement of the two detectors in (a) the same ($x_1 = x_2$) and (b) opposite directions ($x_1 = -x_2$). Solid lines are the theoretical dependencies. Accidental coincidences (~1.5 coincidences per 300 sec) are subtracted. Error bars show standard deviations.

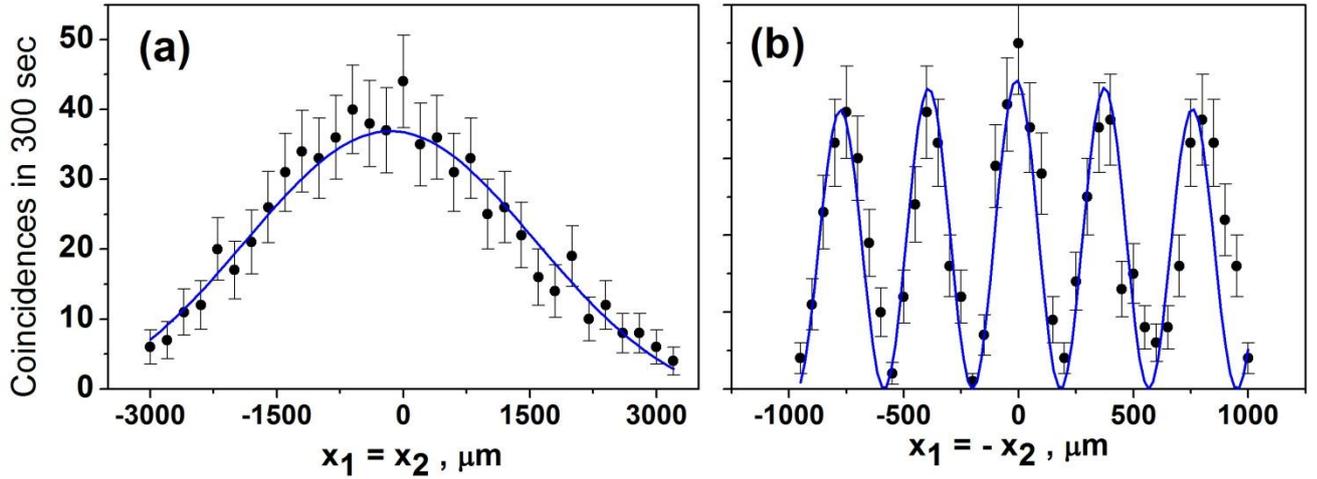

**Figure 6** (Color online) Dependence of the coincidence count rate for separable state on the relative displacement of the two detectors in (a) the same ($x_1 = x_2$) and (b) opposite directions ($x_1 = -x_2$). Solid lines are the theoretical dependencies. Accidental coincidences coincidences (~1.5 coincidences per 300 sec) are subtracted. Error bars show standard deviations.

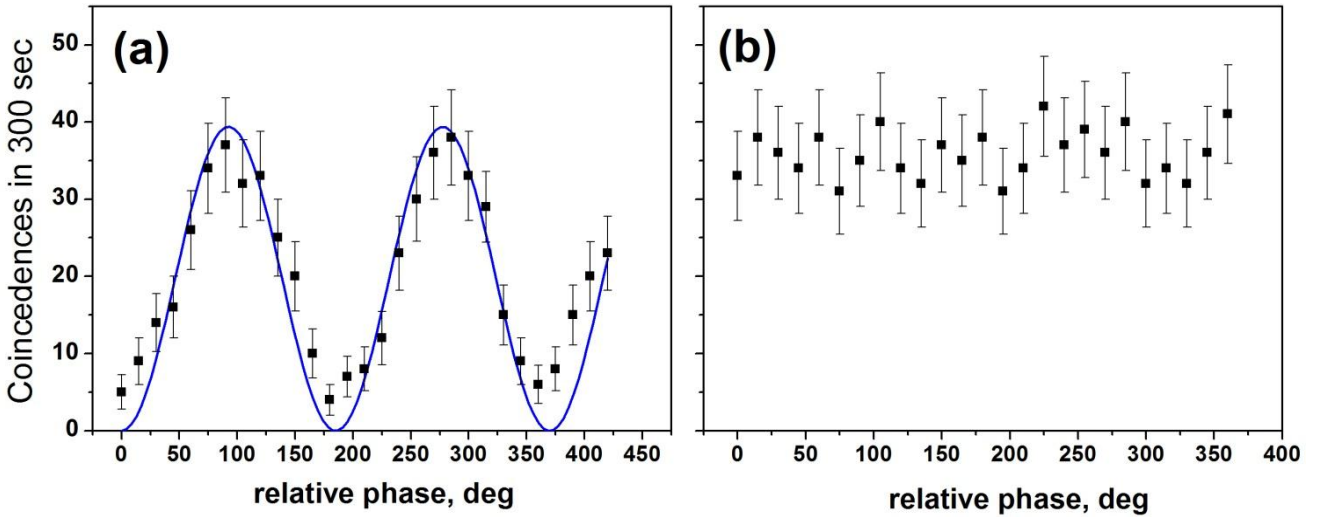

**Figure 7** (Color online) Dependence of the coincidence count rate on the phase detuning of the interferometer for the *NOON*-state (a), and the separable state (b). Two detectors are positioned at $x_1 = x_2 = 0$. Such arrangement corresponds to observation of the coincidences by a single pixel of a 2-photon absorption array. Accidental coincidences (~1.5 coincidences per 300 sec) are subtracted. Error bars show standard deviations. Solid line in (a) is a theoretical dependence, given by Eq.7(c).